\documentclass[showkeys,amsmath,amssymb,a4paper,floatfix,superbib,prb]{revtex4}

\usepackage{graphicx}
\usepackage{natbib}

\begin{document}

\title{A trilinear method for finding null points in a 3D vector space}
\author{A.L. Haynes}
\email{andrew@mcs.st-andrews.ac.uk}
\author{C.E. Parnell}
\affiliation{Department of Mathematics and Statistics, University of St
  Andrews, North Haugh, St Andrews, KY16 9SS, UK} 
 
\begin{abstract}
  Null points are important locations in vector fields, such as a magnetic
  field.  A new technique (a trilinear method for finding null points) is
  presented for finding null points over a large grid of points, such as those
  derived from a numerical experiment.  The method was designed so that the
  null points found would agree with any fieldlines traced using the commonly
  used trilinear interpolation.  It is split into three parts: reduction,
  analysis and positioning, which, when combined, provide an efficient means of
  locating null points to a user-defined sub-grid accuracy.  We compare the
  results of the trilinear method with that of a method based on the Poincar\'e
  index, and discuss the accuracy and limitations of both methods.
\end{abstract}


\maketitle

\section{Introduction} \label{sec:nulls_intro}

A null point (or neutral point, 3D root) is a location where the strength of a
continuous vector field, such as a magnetic field ($\vec{B}$), is locally zero
(i.e.\ \(\vec{B} = \vec{0}\)). Null points in magnetic fields are important
sites for current sheet formation\cite{green65, syrovatsky81, pontin07}, energy
dissipation via either magnetic reconnection\cite{priest96, pontin04, pontin05}
or wave dissipation\cite{mclaughlin05, mclaughlin06b}. Furthermore, null points
are useful for finding other important topological features\cite{bungey96,
parnell96, priest97, schrijver02, cai06}, for example separatrix surfaces and
separators.  Separatrix surfaces (also known as $\Sigma$ or fan surfaces) are a
set of fieldlines which originate or terminate at a null and divide two
topologically distinct regions.  Two separatrix surfaces intersect at a
separator (or $\gamma$ line), a special fieldline which connects two null
points.  Both separatrix surfaces and separators are locations where
reconnection preferentially occurs.  Examples of magnetic null points have been
detected in the laboratory\cite{stenzel02} and in the Earth's
magnetosphere\cite{xiao06} and they are believed to be important in many
astrophysical phenomena such as solar flares and magnetic sub-storms. 

In a finite volume, null points may be found in one of two locations: on the
boundary or inside the volume.  In general, boundary null points may be found
using a 2D method on two of the three components, with the third component used
to check whether any resulting locations are really null points.  Internal null
points may exist anywhere within the volume and require a 3D method to locate
them.  Methods using the Poincar\'e index such as Greene's\cite{greene88b} work
on analytical or numerical fields that are divided into a series of cells. By
considering the cell translated in magnetic field space and then mapped onto
the unit sphere, their methods deduce if the magnetic field space surrounds a
null point. This method can give false positives and false negatives, as 
discussed in detail later. Furthermore, the null point itself is not actually
located, only the grid-cell it is in, unless further assumptions are made, such
as multiple iterations for analytical fields or an interpolation approach is
chosen for a numerical field.  More recently, Zhao\cite{zhao05} has also
proposed essentially the same null finding method.

In recent years the use of large computer simulations of magnetic fields has
increased, and the output of these experiments is usually as a grid of points,
which divide the space into a grid of cells.  To analyse this data a fast and
accurate null-finding method is required.  For fieldlines to be traced within
the experimental domain knowledge of the field between the grid points is
required.  To estimate the field between gridpoints, the field is interpolated,
usually using the method of trilinear interpolation.  Although in general this
does not conserve \(\nabla.\vec{B} = 0\), it is a good first approximation of
the magnetic field between grid points.  For many situations (e.g.\ to
calculate separatrix surface fieldlines), it is necessary to know exactly where
the null points lie to sub-grid resolution.

In this paper, we present a new simple trilinear (TL) method of finding null
points that finds the null points of the actual field traced (using trilinear
interpolation) to subgrid resolution.  We compare our TL method with the
commonly used Greene's method identifying the limitations of each method. 

First, in Section~\ref{sec:nulls_interpolation} we present the linear, bilinear
and trilinear equations.  Then in Section~\ref{sec:nulls_method} the trilinear
(TL) method and Greene's method are described.
Section~\ref{sec:nulls_examples} gives comparative examples of the use of the
two methods and Section~\ref{sec:nulls_comparison} describes the limitations of
the two methods for linear and non-linear fields.  The conclusions
(Section~\ref{sec:nulls_conclusions}) gives a brief discussion on the
advantages and disadvantages of each method.

\section{Interpolation} \label{sec:nulls_interpolation}

\subsection{Linear, Bilinear and Trilinear Equations} 
\label{sec:nulls_inter_linear}

The simplest form of interpolation between $f(0)$ and $f(1)$ in 1D is linear
interpolation, which generates an equation of the form:
\[
  f(x)=f(0)+(f(1)-f(0))x,
\]
For simplicity we write
\[
  f(x)=(1-x)f_{0}+xf_{1}
\]
where \(f_{0}=f(0)\) and \(f_{1}=f(1)\).

We expand this to 2D by interpolating linearly along the horizontal sides of a
square (of unit length) and then by linear interpolation between the two
resulting points.  Points along the bottom of the square are given by
\(p_{0}(x) = (1-x)f_{00}+xf_{10}\) and along the top by \(p_{1}(x) =
(1-x)f_{01}+xf_{11}\) (where \(f_{00}=f(0,0)\) etc.)  Then the field at a point
$(x,y)$ is given by interpolating along the vertical line of constant $x$
between points $p_{0}(x)$ and $p_{1}(x)$.  Hence,
\begin{eqnarray*}
  f(x,y) & = & (1-y)p_{0}(x)+yp_{1}(x) \\
         & = & (1-x)(1-y)f_{00}+x(1-y)f_{10} + (1-x)yf_{01}+xyf_{11} \\
         & = & a+bx+cy+dxy
\end{eqnarray*}
for constants \(a=f_{00}\), \(b=f_{10}-f_{00}\), \(c=f_{01}-f_{00}\) and
\(d=f_{11}-f_{10}-f_{01}+f_{00}\).  Clearly, this gives exactly the same answer
if one interpolates first along the vertical edges of the square for constant
$x$ and then horizontally for constant $y$.

Expanding this to 3D, we obtain the trilinear equation for the field at a point
$(x,y,z)$ inside a cube of unit length,
\begin{eqnarray}
  f(x,y,z) & = & a + bx + cy + dxy + ez + fxz + gyz + hxyz.
\end{eqnarray}
The values of the constants are unique for each cube, and are found to be:
\[\begin{array}{rcl@{\hspace{.35in}}rcl}
  a & = & f_{000}, &
  b & = & f_{100} - f_{000}, \\
  c & = & f_{010} - f_{000}, &
  d & = & f_{110} - f_{100} - f_{010} + f_{000}, \\
  e & = & f_{001} - f_{000}, &
  f & = & f_{101} - f_{100} - f_{001} + f_{000}, \\
  g & = & f_{011} - f_{010} - f_{001} + f_{000}, &
  h & = & f_{111} - f_{110} - f_{101} - f_{011} 
        + f_{100} + f_{010} + f_{001} - f_{000},
\end{array}\]
where \(f_{000} = f(0,0,0)\), etc.

These equations form a simple method of interpolation in up to three
dimensions and have the property that the values on the shared surface of
adjacent cubes are equal.  Higher-order methods such as cubic, bicubic and
tricubic, are not discussed in this paper.

\subsection{Roots of a pair of bilinear equations.}
\label{sub:nulls_inter_bisolve}

We will make use of the roots of a pair of bilinear equations later and so
discuss how to find their roots here.  

For any pair of general bilinear equations 
\begin{equation}
  f_{i}(x,y) = a_{i}+b_{i}x+c_{i}y+d_{i}xy
\end{equation}
with \(i\in\{1, 2\}\) the values of $x$ and $y$ which satisfy
\(f_{1}(x,y)=f_{2}(x,y)=0\) may be found by solving either of the respective
quadratic equations 
\begin{equation}
  \left|\begin{array}{cc} a_{1} & a_{2} \\ c_{1} & c_{2}\end{array}\right| 
  +\left(\left|\begin{array}{cc} a_{1} & a_{2} \\ d_{1} & d_{2}\end{array} 
  \right| + \left|\begin{array}{cc} b_{1} & b_{2} \\ c_{1} & c_{2}\end{array}
  \right|\right)x + \left|\begin{array}{cc} b_{1} & b_{2} \\ d_{1} &
  d_{2}\end{array}\right|x^{2} = 0, 
\end{equation}
or 
\[
  \left|\begin{array}{cc} a_{1} & a_{2} \\ b_{1} & b_{2}\end{array}\right| 
  +\left(\left|\begin{array}{cc} a_{1} & a_{2} \\ d_{1} & d_{2}\end{array} 
  \right| - \left|\begin{array}{cc} b_{1} & b_{2} \\ c_{1} & c_{2}\end{array}
  \right|\right)y + \left|\begin{array}{cc} c_{1} & c_{2} \\ d_{1} &
  d_{2}\end{array}\right|y^{2} = 0.
\]
The resulting 2D coordinate is found using
\begin{equation}
  y = -\frac{a_{i}+b_{i}x}{c_{i}+d_{i}x},
\end{equation}
for any known $x$ where \(i\in\{1,2\}\), or using
\[
  x = -\frac{a_{i}+c_{i}y}{b_{i}+d_{i}y},
\]
for any known $y$ where \(i\in\{1,2\}\).

\section{Methods for finding null points} \label{sec:nulls_method}

The trilinear (TL) and Greene's null-finding methods are both split into three
distinct parts.  The first part, which is the common to both methods, quickly
scans over every grid-cell removing most cells which do not contain a null
point.  The second parts are specific to each method and deduce if a null point
does or does not exists inside a flagged grid-cell.  The third parts locate the
null point within the grid-cell.  All three parts of both methods are described
below.

\subsection*{Reduction} \label{sub:nulls_meth_reduction}

The first stage of both algorithms is to take every grid-cell and use a simple
test to examine whether a null point can exist inside the grid-cell or not.
The test assumes that within each cell the field is linear or trilinear. Thus
an implicit assumption for both methods is there is adequate resolution in the
data.  A direct consequence of the trilinear assumption is that it forces the
range of values of $B_{x}$ inside the cell to lie between the minimum and
maximum values of $B_{x}$ at the corners, and similarly for $B_{y}$ and
$B_{z}$.  Hence, $B_{x}$ can never be zero inside the cell provided that
$B_{x}$ is non-zero and of the same sign at each corner of the cell.

At each corner of the cell, the signs of the $B_{x}$, $B_{y}$ and $B_{z}$ are
considered in turn.  Should any of the three magnetic field components have the
same sign at all eight corners of any cell, then that cell is removed from
further analysis, as it cannot contain a null point.

\subsection{Trilinear Method} \label{sub:nulls_meth_analysis}

The analysis part of the TL method is based upon the fact that a null
point, if it exists, must lie on all three of the following curves:  
\(B_{x}=B_{y}=0\), \(B_{x}=B_{z}=0\) and \(B_{y}=B_{z}=0\).  These curves must
be one of two types:  (i) a circuit inside a cell or (ii) a curve that
extends through the boundary of the cell at either end.  The first case equates
to having two null points within a cell.  This implies considerable sub-grid
structure and clearly shows insufficient grid resolution, as does
Greene\cite{greene88b}.  The second case is the most important, and the TL
algorithm is designed to detect these types of null points.

On the surface of the cell the lines \(B_{x}=0\), \(B_{y}=0\) and
\(B_{z}=0\) are found.  Clearly, due to the trilinear nature inside the
cell, $B_{x}$, $B_{y}$ and $B_{z}$ are bilinear on each face of the
cell.  This allows us to use the analytical solution presented in
Section~\ref{sub:nulls_inter_bisolve}, for any pair of bilinear equations, to
find all intersections of each of the three pairs of lines $B_{x}=0$, $B_{y}=0$
and $B_{z}=0$ on all faces of the cell.  These intersections represent all
locations where any of the intersection curves (e.g.\ $B_{x}=B_{y}=0$) cross
the boundary of the cell.

In general, each of these intersection curves must pierce the boundary of
the cell in pairs.  Indeed, this is a necessary condition for a single
null point to exist along any such intersection curve.  However, it is possible
that there may be more than one pair of points where the intersection curves
cross the cell boundary.  A null point only exists if given any single
intersection curve inside the boundary, with a pair of end points on the
boundary, the third component is of opposite sign at the end points.

Once the existence of a null point has been deduced within a given cell, we
then locate its position to subgrid resolution.  There are many possible
methods to do this.  We have found the following Newton-Raphson method fast and
generally successful.

A 3D version of the iterative Newton-Raphson method for finding roots of
equations has the step:
\begin{equation}
  \vec{x_{n+1}} = \vec{x_{n}} - \left[\left.\nabla\vec{B}\right|_{\vec{x_{n}}}
    \right]^{-1}\vec{B}(\vec{x_{n}}) 
\end{equation}
where \[\left[\nabla \vec{B}\right]_{ij} = \frac{\partial B_{i}}{\partial
  x_{j}}. 
\]
This is repeated until \(\left|\vec{x_{n+1}}-\vec{x_{n}}\right|\) or
\(|\vec{B}(\vec{x_{n}})|\) is less than a given tolerance.  When the trilinear
assumption is used, the differentials can be explicitly written in terms of
the components of $\vec{x}$, and we choose $\vec{x_{0}}$ to be at either the
centre of the cell or a cell corner.  Various starting points are tried
until the iterative method succeeds at some point within the cell.  If
this method fails we split the grid-cell into eight subgrid-cells using
trilinear interpolation and use the TL method again on these eight new cells.

As an aside it is clear that this method may also be implemented for any 2D
field (say when $B_{z}=0$). By using the fast scanning method described in the
first part of the algorithm on the four points around a grid-square 
for $B_{x}$ and $B_{y}$ only, we remove many locations where a 2D null point
cannot exist.  The locations of any null points in this 2D field is then found
by solving the pair of bilinear equations (in $B_{x}$ and $B_{y}$) derived from
the values of the magnetic field at the corners.

\subsection{Greene's method} \label{sub:nulls_meth_greene}

Greene\cite{greene88b} proposed a method based on the Poincar\'e index theorem
to determine the existence of a null point within a given cell.  We give a
brief description here for completeness.

The analysis part of the algorithm, determines the existence of a null point
within a grid-cell.  Each of the six rectangular faces which make up the
boundary of the cell is divided into two triangles, of which there are two
choices (a point which shall be important later).  For each triangle, the
positions of the three magnetic field vectors $\vec{B}_{1}$, $\vec{B}_{2}$ and
$\vec{B}_{3}$ are ordered in a right-handed manner about the normal vector of
the cell.  From this the \emph{area contribution} ($A$) of the triangle, is
calculated from
\begin{equation}
  \tan^{2}\frac{A}{4}=\tan\frac{\theta_{1}+\theta_{2}+\theta_{3}}{4}
    \tan\frac{\theta_{1}+\theta_{2}-\theta_{3}}{4}
    \tan\frac{\theta_{2}+\theta_{3}-\theta_{1}}{4} 
    \tan\frac{\theta_{3}+\theta_{1}-\theta_{2}}{4},
\end{equation}
and
\[\begin{array}{rcl@{\hspace{.35in}}rcl@{\hspace{.35in}}rcl}
  \cos\theta_{1} & = & \displaystyle 
    \frac{\vec{B}_{2}.\vec{B}_{3}}{|\vec{B}_{2}| |\vec{B}_{3}|}, & 
  \cos\theta_{2} & = & \displaystyle
    \frac{\vec{B}_{1}.\vec{B}_{3}}{|\vec{B}_{1}| |\vec{B}_{3}|}, & 
  \cos\theta_{3} & = & \displaystyle
    \frac{\vec{B}_{1}.\vec{B}_{2}}{|\vec{B}_{1}| |\vec{B}_{2}|}. 
\end{array}\]
Finally the area contribution, $A$, is chosen to be negative if
\(\vec{B}_{1}.\vec{B}_{2} \times \vec{B}_{3}\) is negative, else the area
contribution is taken to be positive.  The deviation of these area
contributions is trivial to show using spherical geometry techniques.

The topological degree of the cell is determined by summing up all twelve area
contributions, and then dividing the result by $4\pi$.  The cell contains a
null point if the topological degree is non-zero (a positive null if the degree
is $-1$, a negative null if $+1$).

Further refinement is given by bisecting the cell to gain greater accuracy.
Greene\cite{greene88b} also implemented a method of guessing the location of a
null point near the point \(\vec{x}_{0} = \langle x_{0}, y_{0}, z_{0} \rangle\)
(the secant method) by solving the three simultaneous equations
\begin{equation}
  \frac{\vec{B}_{x}-\vec{B}_{0}}{\delta
  x}(x-x_{0})+\frac{\vec{B}_{y}-\vec{B}_{0}}{\delta y}(y-y_{0})
  \frac{\vec{B}_{z}-\vec{B}_{0}}{\delta z}(z-z_{0}) = -\vec{B}_{0},
\end{equation}
for $x$, $y$ and $z$, where \(\vec{B}_{0}=\vec{B}(\vec{x}_{0})\),
\(\vec{B}_{x}=\vec{B}(\vec{x}_{0}+\delta x\,\hat{\vec{x}})\),
\(\vec{B}_{y}=\vec{B}(\vec{x}_{0}+\delta y\,\hat{\vec{y}})\) and
\(\vec{B}_{z}=\vec{B}(\vec{x}_{0}+\delta z\,\hat{\vec{z}})\), and $\delta x$,
$\delta y$ and $\delta z$ are small.  This approach is similar to the
Newton-Raphson method which was described earlier.

\section{Examples comparing methods} \label{sec:nulls_examples}

To demonstrate the differences between the TL and Greene's method, we consider
a couple of examples.

\subsection{Example I} \label{sub:nulls_examples_michelle}

\begin{figure}
  \begin{center}
  \includegraphics{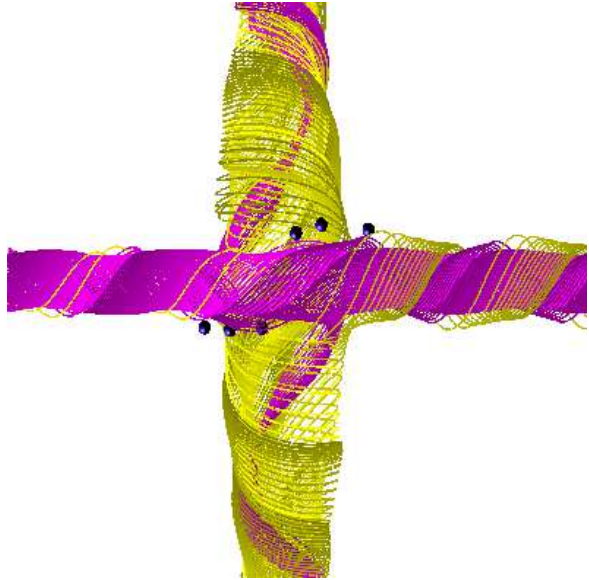}
  \end{center}
  \caption{Example showing the null points found by both null-finding methods
    on a sample frame of a numerical experiment\cite{murray06} involving two
    twisted flux tubes passing through one another.  Both methods found six
    null points (shown as spheres) between the two flux tubes, which are shown
    as sets of fieldlines traced from the sides of the box.}
  \label{fig:nulls_example_michelle}
\end{figure}

The speed and accuracy of the null-finding methods is tested on a frame of a
numerical experiment involving a twisted flux tube rising through another
twisted flux tube beneath the solar photosphere\cite{murray06}.  Both the TL
method and Greene's method positively identified the same six grid-cells
containing null points in the frame (see Figure~%
\ref{fig:nulls_example_michelle}).

The TL method took just 24.9 seconds to do this compared to Greene's 25.0
seconds, hence the algorithms operate at comparable speeds.   The reduction
phase reduced the number of grid-cells to search from 581433 to 781 (a
reduction of 99.87\%).

\subsection{Example II} \label{sub:nulls_example_john}

\begin{figure}
  \begin{center}
  \includegraphics{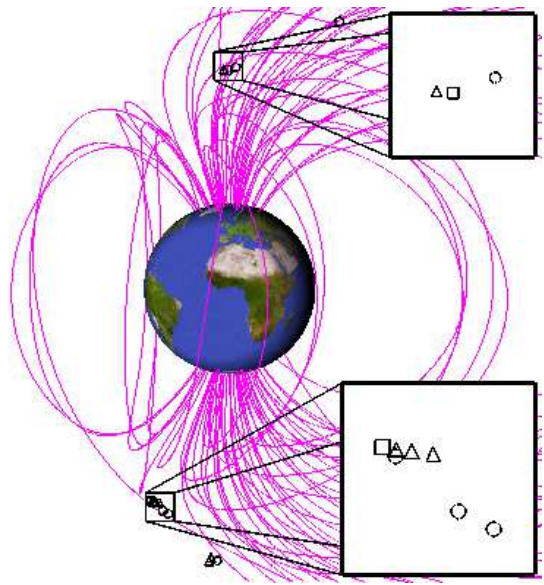}
  \end{center}
  \caption{A sample frame of the Earth's magnetospheric magnetic
    field\cite{dorelli07} used to compare the null finding codes.  The small
    symbols represent points identified as nulls (triangle -- TL method, square
    -- Greene's test, circle -- both TL method and Greene's test).  Note that
    two distant null points found by both methods are not shown.  The lines
    represent sample fieldlines traced from the poles of the Earth.}
  \label{fig:nulls_example_dorelli}
\end{figure}

The accuracy of the TL method and Greene's test were compared using a sample
frame of a simulation of the Earth's magnetosphere\cite{dorelli07}.  Both
methods returned the same eight null points, but Greene's test found another
two possible null points and the TL method found a further six (see Figure~%
\ref{fig:nulls_example_dorelli}).  Clearly in a numerical experiment it is not
possible to know exactly what the field inside the cells without rerunning the
experiment at higher resolution, which is rarely practical.  However, if we
assume the field to be trilinear within each cell, it is possible to show that
Greene's extra ``nulls'' are false positives and that TL's extra nulls are
actual null points.  In the next section, we shall discuss the accuracy of both
techniques, with respect to analytic types of field.

\section{Discussion of accuracy of methods} \label{sec:nulls_comparison}

\subsection{Linear Fields}

All linear fields can be written in the form
\begin{equation}
  \vec{B}(\vec{x}) = \vec{a}x + \vec{b}y + \vec{c}z + \vec{d} =
  \left[\begin{array}{ccc} \vec{a} & \vec{b} & \vec{c}
  \end{array}\right]\vec{x} + \vec{d}, 
\end{equation}
where $\vec{a}$, $\vec{b}$, $\vec{c}$ and $\vec{d}$ are arbitrary constant
vectors of the form \(\vec{a} = \langle a_{1}, a_{2}, a_{3} \rangle\).
Provided the matrix \([\nabla \vec{B}] = [\vec{a}\;\;\vec{b} \;\;\vec{c}]\) is
non-singular (which is the case for a generic 3D field), this can be rewritten
as
\(
  \vec{B}(\vec{x}) = [\nabla \vec{B}](\vec{x}-\vec{x}_{N}),
\)
where \(\vec{x}_{N} = -[\nabla\vec{B}]^{^{-1}}\vec{d}\).

\subsubsection{Greene's Method}

To demonstrate that Greene's method will always correctly identify null points
in a linear field, we need only consider the signs of the triangular area
contributions ($A$), which are each given by the triple scalar product
\(\chi(\triangle PQR) = \vec{B}(\vec{P}) \cdot \vec{B}(\vec{Q}) \times
\vec{B}(\vec{R})\) for triangle $\triangle PQR$.  Hence,
\begin{eqnarray}
  \chi(\triangle PQR) 
  & = & \vec{B}(\vec{P}) \cdot \vec{B}(\vec{Q}) \times \vec{B}(\vec{R})
        \nonumber \\ 
  & = & \left\{\left[\nabla\vec{B}\right](\vec{P}-\vec{x}_{N})\right\}
        \cdot\left\{\left[\nabla\vec{B}\right](\vec{Q}-\vec{x}_{N})\right\}
        \times\left\{\left[\nabla\vec{B}\right](\vec{R}-\vec{x}_{N})\right\}
        \nonumber \\ 
  & = &  \left|\nabla\vec{B}\right| 
        \left(\left\{ \vec{P}-\vec{x}_{N} \right\} \cdot  \left\{
        \vec{Q}-\vec{x}_{N} \right\} \times \left\{ \vec{R}-\vec{x}_{N}
        \right\} \right)  
\end{eqnarray}
Clearly \(|\nabla\vec{B}|\) is constant and hence of fixed sign throughout the
domain.  So if we determine $\chi$ for all triangles in a right-handed manner
relative to the outwards normal vector of the volume enclosed by the triangles
then, if there is a null point inside this volume, 
the sign for each  $\chi$ will be the same.  If positive there is a negative
null point inside the volume, if negative a positive null point.  In
particular, note that the sign of both triangles on any one face of a cell will
be the same.  Hence, Greene's method will always find a null point if one
exists within the volume.  If the null point is outside the cell, the triangles
relative to $\vec{x}_{N}$ have $\chi$ of mixed sign.

Clearly, if a null point exists within a cell then the topological degree is
either $\pm 1$.  For any linear field containing a null point, it is possible
to choose cells of any size and still resolve the field within the cell.  This
lets us consider a large cell that contains a null point.  This large cell may
be divided in to a grid of $n$ smaller cells, of which $n-1$ do not contain a
null point.  The topological degree of this big cell must be the sum of the
topological degrees of the smaller cells which span it.  Furthermore, we note
that the topological degrees for the large cell and the small cell containing
the null point are 1 (assuming we have a negative null point, without loss of
generality).  Thus, by considering the case of $n=2$ and working up, it is
clear that the topological degree of all smaller cells without a null point may
be deduced to be zero.  Therefore, the topological degree determined by
Greene's method will always be correct for any cell in a linear field.

\subsubsection{Trilinear Method}

Similarly we can show that the TL method also always correctly identifies a
null point in a linear field.  If we consider the surface 
\( 
  \left\langle a_{1}, b_{1}, c_{1} \right\rangle.\vec{x} = -d_{1},
\)
where $B_{x} = 0$, and the surface
\(
  \left\langle a_{2}, b_{2}, c_{2} \right\rangle.\vec{x} = -d_{2},
\)
where $B_{y} = 0$, then their intersection will be a straight line in the
direction
\(
  \vec{n} = \left\langle a_{1}, b_{1}, c_{1} \right\rangle \times \left\langle
    a_{2}, b_{2}, c_{2} \right\rangle,
\)
and the line will be of the form 
\(
  \vec{x}(s) = \vec{x}_{0} + \vec{n}s,
\)
where $\vec{x}_{0}$ is any point on both planes.  An obvious choice for this
point is the null point, $\vec{x}_{N}$, hence
\(
  \vec{x}(s) = \vec{n}s - \left[\begin{array}{ccc} \vec{a} & \vec{b} & \vec{c}
    \end{array}\right]^{-1}\vec{d}. 
\)
Calculating $\vec{B}$ along this line, we find
\begin{eqnarray}
  \vec{B}(s) & = & \left[\begin{array}{ccc} \vec{a} & \vec{b} & \vec{c}
                   \end{array}\right] \left\{ \vec{n}s -
                   \left[\begin{array}{ccc} \vec{a} & \vec{b} & \vec{c} 
                   \end{array}\right]^{-1}\vec{d}\right\} + \vec{d}
		   \nonumber \\
             & = & \left[\begin{array}{ccc} \vec{a} & \vec{b} & \vec{c}
                   \end{array}\right]\left(\left\langle a_{1}, b_{1}, c_{1}
		   \right\rangle \times \left\langle a_{2}, b_{2}, c_{2}
		   \right\rangle s\right) - \vec{d} + \vec{d} \nonumber \\
             & = & \left|\left[\begin{array}{ccc} \vec{a} & \vec{b} & \vec{c}
                   \end{array}\right]\right| s \hat{\vec{z}}. 
\end{eqnarray}
From this it is obvious that $B_{z}$ does not change sign along its length
except at the null point (when $s = 0$).

We have a two cases:
\begin{enumerate}
  \item {Part of the line is inside the cell:}  Since the line is
     straight and of infinite length, then for any finite volume which the line
     passes through, the line intersects the surface at least twice.  The
     TL method detects these locations.  Since the sign of $B_{z}$ changes
     only at the single null on this line, the null point must be inside the
     cell if, and only if, the sign of $B_{z}$ is different at the locations
     where it intersects the surface of the cell.  Thus this agrees with the TL
     method.
  \item {None of the line is inside the cell:}  Since the (only) null
    point is on this line, it cannot be inside the cell.  As the TL method
    does not detect this line on the surface of the cell, it rightly returns
    that no null exists within the cell. 
\end{enumerate}

This above argument holds for all pairings of field components.  Furthermore,
since the field is linear, the final positioning part will succeed after the
first step, as the iterative step,
\begin{equation}
  \vec{x}_{i+1} 
    =   \vec{x}_{i} - \left[\nabla\vec{B}\right]^{-1} \vec{B}(\vec{x}_{i})
    =   \vec{x}_{i} - \left[\begin{array}{ccc} \vec{a} & \vec{b} & \vec{c}
        \end{array}\right]^{-1} \left(\left[\begin{array}{ccc}
           \vec{a} & \vec{b} & \vec{c} 
        \end{array} \right] \vec{x}_{i} + \vec{d} \right)
    =   - \left[\begin{array}{ccc} \vec{a} & \vec{b}
         & \vec{c} \end{array}\right]^{-1}\vec{d} 
    =   \vec{x}_{N},
\end{equation}
for any value of $\vec{x}_{i}$.

\subsection{Nonlinear Fields}

Nonlinear fields are clearly more complicated than linear fields and this is
where the results from the two methods can diverge.

\subsubsection{Greene's Method}

We return to the results of Example~II, where we have noted discrepancies
between the results of Greene's method and the TL method.  By swapping the
choice of triangles for each face of the grid-cells for Greene's method, we
find fourteen null points with this alternative implementation. Six are new,
four the same as those found previously by the TL method alone and a further
four are the same as four found previously by both the TL method and the
original Greene's implementation.  This leads to a curious result about the
nature of Greene's method when implemented for non-linear fields.

It is known that the sum of all the area contributions in Greene's method must
sum up to an integer, and the triangles chosen for each face may be changed
independently from the other faces.  If the result of Greene's algorithm
changes depending on the choice of triangles, then the difference between the
area contributions using one choice of triangles on the face must be an integer
difference to the other choices.  Using Example~II, we found that different
implementations of Greene's method could add or remove a null point, or even
(in a few cases) change the sign of the null point.

To further this investigation, we analysed a set of trilinear fields where the
solenoidal condition (\(\nabla.\vec{B} = 0\)) was satisfied.  This choice of
non-linear field is chosen because the field inside numerical cells is
typically assumed to be trilinear and because the TL method can easily find
these null points.  Numerous fields were found (using a method described 
in Appendix~\ref{app:greene_failure}) where at least one of the faces of a cell
would change the topological degree in the cell depending on the choice of
triangles producing both false negatives and false positives.  Although most of
the cases involved the null point existing near the face in question, some
cases were found where null points were near the grid-cell's centre.

An example trilinear field is given below which contains two null points and
demonstrates the curious behaviour of Greene's method under different
implementations.
\begin{equation}
  \vec{B}(\vec{x}) = \left(\begin{array}{c}
    -1.80 \\  0.44 \\ -0.67
  \end{array}\right)+\left(\begin{array}{ccc}
    -2.57 &   6.92 &   0.44 \\
    -3.05 &   2.09 &   0.20 \\
    -2.30 &   8.69 &   0.48 \\
  \end{array}\right)\!\!\left(\begin{array}{c} 
     x    \\  y    \\  z 
  \end{array}\right) + \left(\begin{array}{ccc}
     0.02 &   0.46 &  -1.40 \\
    -0.46 &  -0.34 &   1.46 \\
     1.40 &  -1.46 &  -8.29
  \end{array}\right)\!\!\left(\begin{array}{c}
     yz   \\  xz   \\  xy
  \end{array}\right).
  \label{eq:nulls_tri_example}
\end{equation}

\begin{figure}
  \begin{center}
    \resizebox{5in}{!}{\includegraphics{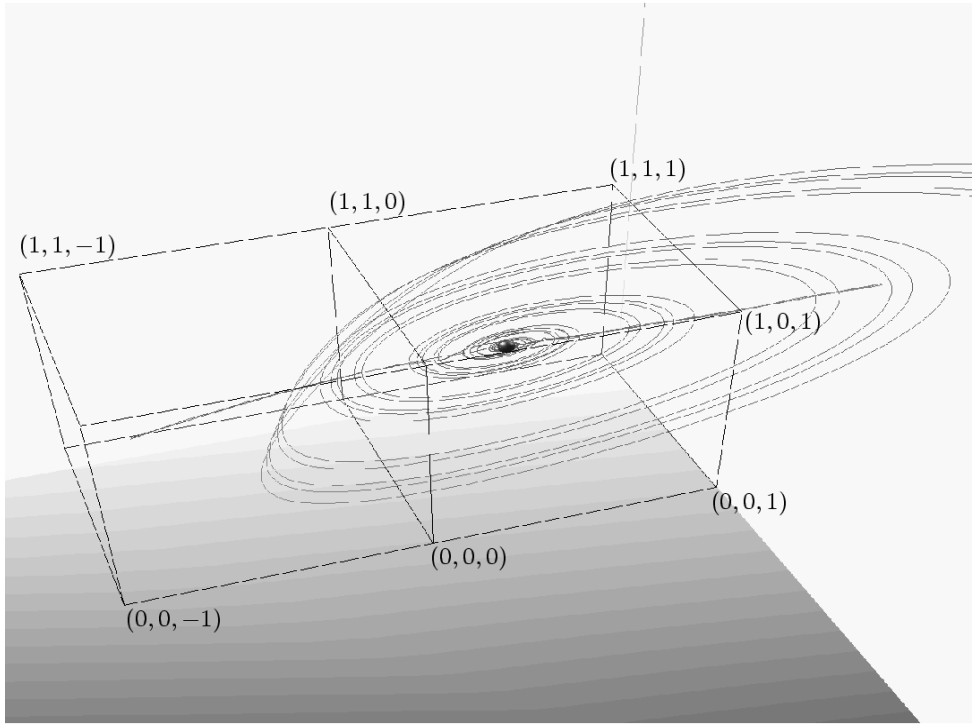}}
  \end{center}
  \caption{Fieldlines showing the spines and separatrix surfaces of the null
    point in the trilinear field given by Equation~\ref{eq:nulls_tri_example}
    where Greene\cite{greene88b} may fail upon implementation.  The $z=0$ face
    which changes the topological degree in the left cell from -1 to 0 is the
    one between the two cells drawn.}
  \label{fig:nulls_tri_example}
\end{figure}

In this example field, we have a negative null point at \((0.59, 0.49, 0.44)\)
and a positive null point at \((0.23, 0.11, 3.23)\).  In the cell \([0,1]^2\!
\times\! [-1,0]\) which contains no null point, Greene's algorithm gives the
topological degree to be $-1$ or $0$ depending on how the face $z=0$ is divided
(see Figure~\ref{fig:nulls_tri_example}).  We now extend our region of interest
to the sequential group of cells \([0,1]^{2}\! \times\! [-1,0]\),
\([0,1]^{3}\), \([0,1]^{2}\! \times\! [1,2]\), \([0,1]^{2}\! \times\! [2,3]\)
and \([0,1]^{2}\! \times\! [3,4]\).  In this case we have four internal faces
between the cells.  If we consider only implementations of Greene's method
where the triangles coincide on these faces, of which there are sixteen, each
yields a different result: one with no null points in any cell (i.e.\
topological degree of the cells is $(0,0,0,0,0)$ for increasing $z$), ten
results detect one positive and one negative null point (e.g.\ $(-1,1,0,0,0)$,
$(-1,0,1,0,0)$, $(0,0,-1,0,1)$ where order must be $-1$ then $1$ ignoring zeros
-- note one of these is the correct solution $(0,-1,0,0,1)$) and five with two
positive and two negative null points (e.g.\ $(-1,1,0,-1,1)$, again in order
$-1$, $1$, $-1$, $1$ with the $0$ in any one of the five cells).  Thus,
Greene's method may (i) move null points a significant distance or (ii) change
the number of null points.  Furthermore it is not possible to say which of the
above implementations of Greene's method is correct without either prior
knowledge of the result or increasing the resolution of the grid.  In an
analytical field case, bisecting over the complete region will generally reduce
the likelihood of such a failure, but on a numerical grid this would require
assumptions about the magnetic field within the domain.

In this example the ratio of the maximum absolute value of the nonlinear terms
to the linear terms is 0.95, showing significant nonlinearity.  But significant
nonlinearity is not essential for Greene's method to give a false reading, as
Example~II includes a number of cases where the ratio in such cases is lower
than 0.25.

Note in the above example we always choose the triangles on the internal cell
faces to coincide for both cells on either side of the face.  Thus in swapping
the choice of triangles (for both grid-cells), we can only add or remove null
points in opposite polarity pairs or move the existing null point between
grid-cells leaving the overall topological degree of the region unchanged.
Should the triangles of only one of these two cells on either side of the face
change, then this cell could create or destroy a single null point.  Hence, it
is important that when implementing Greene's method, the triangles of all
internal faces are chosen to coincide.

\subsubsection{Trilinear Method}

We now discuss the accuracy of the TL method.  First, we shall look at the
nature of trilinear fields, followed by a discussion of the accuracy for a
general non-linear field.

Trilinear fields have four non-linear terms ($xy$, $xz$, $yz$ and $xyz$), and
are solvable using the TL method.  An important point to note though about
trilinear fields is that they are not rotationally preserved, e.g.\ if we take
the function $f(x,y,z)=1+xy$, and rotate it by 45$^{\circ}$ about
$\hat{\vec{z}}$ then we obtain the function
\[
  F(X,Y,Z) = f\left(\frac{X-Y}{\sqrt{2}}, \frac{X+Y}{\sqrt{2}},Z\right)
           = 1+\frac{X^{2}-Y^{2}}{2},
\]
where $F(X,Y,Z)$ is the function in the rotated space $X,Y,Z$, which is
obviously not trilinear.  Hence, if we rotate a grid by any angle, the exact
results of the TL method may vary.  Since the TL method only locally
approximates the field as trilinear, it may still be used to locate the null
point in the rotated field.

If we relax the conditions to a general non-linear field, it is possible using
the TL method for a null point to be found in a different cell if the field is
sufficiently non-linear.  As a null point exists on the intersection of a line
($B_{x} = B_{y} = 0$, say) and a surface ($B_{z} = 0$, say), and both of these
are either closed or extend through the boundary of the full domain, null
points can only be removed or added in pairs within the domain, or singularly
though the boundary of the full domain.  If we consider that a null point is at
the intersection of the three surfaces ($B_{x} = 0$, $B_{y} = 0$ and $B_{z} =
0$) and hence at the corner of the eight volumes bounded by these three
surfaces.  This null point may only be removed if any of these eight volumes is
removed.  Provided the grid is fine enough, then there should exist a grid
point within each volume bounded by the three surfaces.  Since the grid point
has a known field, it must stay within the same volume, and hence the volume
cannot be destroyed, even if the resulting trilinear field is vastly different
from the real field.  Thus, no null points can disappear for the TL method
(although its position may vary) if this condition is satisfied.  We clarify
this with a 2D example.

\begin{figure}
  \includegraphics{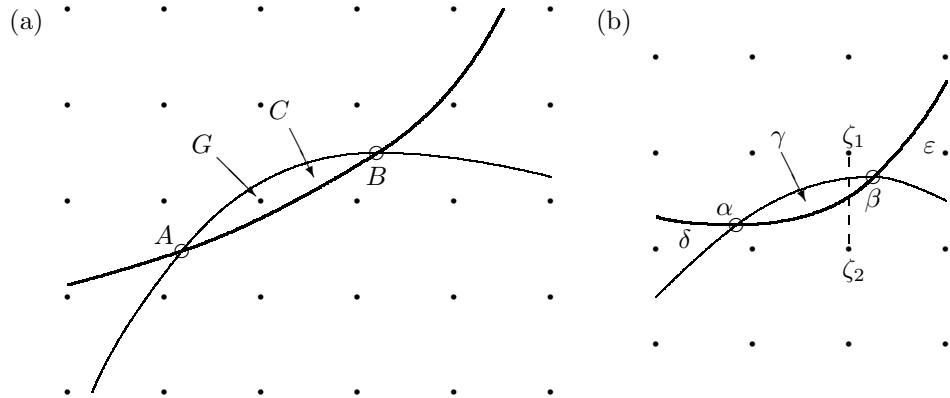}
  \caption{(a) Example configuration of the TL method which cannot lose a
     null point. The thick line denotes $B_{x} = 0$, the thin line $B_{y} = 0$
     and the hollow circles ($A$ and $B$) represent null points.  The dots
     represent the grid used in the TL method.  $C$ refers to the area between
     the null points bounded by the thick and thin lines.
     (b) Example configuration of the TL method which may lose a null point
     for nonlinear fields. The thick line denotes $B_{x} = 0$, the thin line
     $B_{y} = 0$ and the hollow circles ($\alpha$ and $\beta$) represent
     null points.  The dots represent the grid used in the TL method.
     $\gamma$, $\delta$ and $\varepsilon$ refer to the area bounded by the
     thick and thin lines.}
  \label{fig:nulls_rap_success}
  \label{fig:nulls_rap_failure}
\end{figure}

If the above argument is considered in 2D, the surfaces ($B_{x} = 0$, etc.)
become lines and the volumes bounded by the surfaces become areas bounded by
lines.  An example grid with $B_{x} = 0$ and $B_{y} = 0$ is shown in Figure~%
\ref{fig:nulls_rap_success}a.  For a null point to be found in 2D, the TL
method looks for intersections of the lines $B_{x} = 0$ and $B_{y} = 0$.  Since
the TL method assumes the field within four grid points to be bilinear, these
lines may be deformed from the real field.  However, the real field is known at
each grid point, so the lines $B_{x} = 0$, etc.\ found must pass through the
same edge as the real lines $B_{x}=0$, etc., since otherwise the sign of
$B_{x}$, etc.\ at the grid points would change.  Should a grid point exist
within each area bounded by zero lines, for example area $C$ which contains
grid point $G$, (see Figure~\ref{fig:nulls_rap_success}a), the area may be
deformed, but can never be removed by any continuous interpolation of the
field, since grid point ($G$) can only exist within this area.  Hence, the null
points ($A$ and $B$) must always be preserved.  In the case of an area with no
grid point within it, for example area $\gamma$ (see Figure~%
\ref{fig:nulls_rap_failure}b), the area may be deformed and shrunk by a
continuous interpolation of the field until the null points $\alpha$ and
$\beta$ meet and are annihilated, as the areas $\delta$ and $\varepsilon$
merge.  In this case, pairs of null points may be lost (or added by the
converse argument) by the TL method.  In this example, if the field is locally
close to linear or trilinear, the approximation of the intersection of the
lines ($B_{x} = 0$ and  $B_{y} = 0$) to the line between the grid points
$\zeta_{1}$ and $\zeta_{2}$ will be roughly correct and the TL method will
still find the null points even with no grid point in area $\gamma$.  For false
readings in the TL method to occur, the field must be very highly nonlinear,
and hence severely under-resolved.

\section{Conclusion} \label{sec:nulls_conclusions}

It is well known that the location of null points is important in the
understanding of magnetic topology and for some types of reconnection and wave
dissipation.  Here, we have presented a new technique, the trilinear (TL)
method, for finding null points in a magnetic field, in particular for whose
field is calculated using numerical code.  The TL method includes three stages.
A reduction stage to quickly remove most grid-cells which do not contain a null
point.  An analysis to positively identify cells containing a null point using
the bilinear nature of the field on the boundaries, and finally a stage to
determine the exact location of the null point according to trilinear
interpolation.

We compare our TL method with the Poincar\'e index based method of
Greene\cite{greene88b}.  Both the TL and Greene's method
are accurate for all linear fields.  The TL method is in general accurate for
most nonlinear fields.  Null points may be falsely created or destroyed, either
in pairs or lost through the boundary of the domain where the field within a
cell is highly nonlinear.  If we restrict the field to being trilinear within
cells, the TL method is accurate except when two null points exist within one
cell making it highly suitable for numerical magnetic fields. 

Greene's method has similar failings to the TL method.  In Greene's method the
choice of triangles on the cell faces is arbitrary.  In certain (moderately as
well as highly) non-linear fields different implementations of Greene's
algorithm, assuming a different choice of triangles, give rise to failings in
the method.  If triangles on internal faces are kept coincident, null points
can only be lost in pairs from inside the domain (or singularly through the
boundary).  However, if the triangles on internal faces are not kept coincident
then null points may be created and destroyed singularly from anywhere within
the domain.  A single implementation of Greene's method should not be used on
its own as it is not clear how many or where the null points should be.  A
result can only be relied upon if all coincident triangle implementations
agree.  By using different implementations of Greene's method it is possible to
detect where Green's method is failing, but this does not help give the correct
answer unless the resolution of the grid can be increased. However, Greene's
algorithm is very useful for finding null points in analytical fields, whilst
the TL method is most applicable for numerical fields.

\begin{acknowledgments}
  ALH would like to thank the University of St Andrews for his financial
  support during his Ph.D. and CEP acknowledges the support of Particle Physics
  and Astronomy Research Council through an Advanced Fellowship.  We would also
  like to thank John Dorelli and Michelle Murray for providing sample frames
  from their numerical experiments.  We also wish to thank our anonymous
  referee for all the useful comments which have significantly improved this
  paper.
\end{acknowledgments}

\appendix

\section{Failure of Greene's algorithm} \label{app:greene_failure}

A means to finding a cell where Greene's algorithm fails is as follows.
Looking at a face of a cell, say $z=0$ with $x,y\in[0,1]$, the magnetic field
vectors are $\vec{B}_{000}$, $\vec{B}_{100}$, $\vec{B}_{010}$ and
$\vec{B}_{110}$ at the corners.  On our face, we wish to have each pair of
triangular area contributions to have opposite sign.  So, for instance, the
triple scalar products of the magnetic field vectors in the corners may satisfy
\begin{equation}\begin{array}{rcl@{\hspace{.5in}}rcl}
  \vec{B}_{000} . \vec{B}_{100} \times \vec{B}_{110} & > & 0, &
  \vec{B}_{000} . \vec{B}_{110} \times \vec{B}_{010} & > & 0, \\
  \vec{B}_{000} . \vec{B}_{100} \times \vec{B}_{010} & < & 0, &
  \vec{B}_{100} . \vec{B}_{110} \times \vec{B}_{010} & < & 0,
\end{array}\end{equation}
or have all inequalities reversed.  Since multiplying any vector by a positive
constant does not change the sign of the triple vector product, we shall assume
all vectors are normalised.  Let us then choose an invertible matrix $T$, which
satisfies the following three properties:
\begin{equation}\begin{array}{rcl@{\hspace{.5in}}rcl@{\hspace{.5in}}rcl}
  T\vec{x} & = & \vec{B}_{000}, &
  T\left(\begin{array}{c} \cos\theta_{1} \\ \sin\theta_{1} \\ 0 \end{array}
    \right) & = & \vec{B}_{100}, &
  |T| & = & 1 \\
\end{array}\end{equation}
for some $\theta_{1}$.  From this we find a $\theta_{2}$, $\theta_{3}$,
$\phi_{2}$ and $\phi_{3}$ such that 
\begin{equation}\begin{array}{rcl@{\hspace{.5in}}rcl}
  T\left(\begin{array}{c} \cos\theta_{2}\cos\phi_{2} \\
    \sin\theta_{2}\cos\phi_{2} \\ \sin\phi_{2} \end{array}\right) & = &
    \vec{B}_{010}, &
  T\left(\begin{array}{c} \cos\theta_{3}\cos\phi_{3} \\
    \sin\theta_{3}\cos\phi_{3} \\ \sin\phi_{3} \end{array}\right) & = &
    \vec{B}_{110}.
\end{array}\end{equation}
Thus our inequalities become
\[\begin{array}{rcl@{\hspace{.5in}}rcl}
  \left|\begin{array}{ccc}
    1 & \cos\theta_{1} & \cos\theta_{2}\cos\phi_{2} \\
    0 & \sin\theta_{1} & \sin\theta_{2}\cos\phi_{2} \\
    0 & 0              & \sin\phi_{2} 
  \end{array}\right| & > & 0, &
  \left|\begin{array}{ccc}
    1 & \cos\theta_{2}\cos\phi_{2} & \cos\theta_{3}\cos\phi_{3} \\
    0 & \sin\theta_{2}\cos\phi_{2} & \sin\theta_{3}\cos\phi_{3} \\
    0 & \sin\phi_{2}               & \sin\phi_{3} 
  \end{array}\right| & > & 0, \\\\
  \left|\begin{array}{ccc}
    1 & \cos\theta_{1} & \cos\theta_{3}\cos\phi_{3} \\
    0 & \sin\theta_{1} & \sin\theta_{3}\cos\phi_{3} \\
    0 & 0              & \sin\phi_{3}
  \end{array}\right| & < & 0, &
  \left|\begin{array}{ccc}
    \cos\theta_{1} & \cos\theta_{2}\cos\phi_{2} & \cos\theta_{3}\cos\phi_{3} \\
    \sin\theta_{1} & \sin\theta_{2}\cos\phi_{2} & \sin\theta_{3}\cos\phi_{3} \\
    0              & \sin\phi_{2}               & \sin\phi_{3}
  \end{array}\right| & < & 0.
\end{array}\]
From these we deduce that
\begin{equation}\begin{array}{rcl@{\hspace{.5in}}rcl}
  \sin\theta_{1}\sin\phi_{2} & > & 0, &
  \sin\theta_{2}\cos\phi_{2}\sin\phi_{3} & > &
    \sin\theta_{3}\cos\phi_{3}\sin\phi_{2}, \\
  \sin\theta_{1}\sin\phi_{3} & < & 0, &
  \cos\phi_{2}\sin\phi_{3}\left|\begin{array}{cc} \cos\theta_{1} &
    \cos\theta_{2} \\\sin\theta_{1} & \sin\theta_{2} \end{array}\right| & < &
  \sin\phi_{2}\cos\phi_{3}\left|\begin{array}{cc} \cos\theta_{1} &
    \cos\theta_{3} \\\sin\theta_{1} & \sin\theta_{3} \end{array}\right|.
\end{array}\end{equation}
Now, let 
\(
  \alpha = {\tan \phi_{2}}/{\tan\phi_{3}}
\)
so that the last two inequalities become
\begin{equation}\begin{array}{rcl@{\hspace{.5in}}rcl}
  \sin\theta_{2} & > & \alpha\sin\theta_{3}, &
  \sin\left(\theta_{2} - \theta_{1}\right) & < &
    \alpha\sin\left(\theta_{3} - \theta_{1}\right),
\end{array}\end{equation}
provided that \(\cos\phi_{2}\sin\phi_{3} > 0\), otherwise the inequalities are
reversed.  Now we may choose any value for $\alpha$ and $\theta_{1}$, then find
values for $\theta_{2}$ and $\theta_{3}$ so that either pair of the above
equalities hold, and choose a $\Phi_{2}$ and $\Phi_{3}$ which satisfy \(\alpha
= {\tan \Phi_{2}}/{\tan\Phi_{3}}\).  To satisfy the final conditions, let
$\phi_{2}$ and $\phi_{3}$ be given from this table:
\begin{center}\begin{tabular}{c|*{3}{c@{\hspace{1.5em}}}c}
  $\cos\Phi_{2}\sin\Phi_{3}$  & $\sin\theta_{1}\sin\Phi_{2} > 0$ &
    $\sin\theta_{1}\sin\Phi_{2} > 0$ & $\sin\theta_{1}\sin\Phi_{2} < 0$ &
    $\sin\theta_{1}\sin\Phi_{2} < 0$ \\
  & $\sin\theta_{1}\sin\Phi_{3} < 0$ & $\sin\theta_{1}\sin\Phi_{3} > 0$ &
    $\sin\theta_{1}\sin\Phi_{3} < 0$ & $\sin\theta_{1}\sin\Phi_{3} > 0$ \\
    \hline 
  \raisebox{-0.6em}[0em][0em]{$>0$} & $\phi_{2} = \Phi_{2}$ & $\phi_{2} =
    \pi-\Phi_{2}$ & $\phi_{2} = -\Phi_{2}$ & $\phi_{2} = \Phi_{2}-\pi$ \\
  & $\phi_{3} = \Phi_{3}$ & $\phi_{3}= -\Phi_{3}$ &
    $\phi_{3} = \pi-\Phi_{3}$ & $\phi_{3} = \Phi_{3}-\pi$ \\ \hline
  \raisebox{-0.6em}[0em][0em]{$<0$} & $\phi_{2} = \pi-\Phi_{2}$ & $\phi_{2} =
    \Phi_{2}$ & $\phi_{2} = \Phi_{2}-\pi$ &$\phi_{2} = -\Phi_{2}$ \\
  & $\phi_{3} = \pi-\Phi_{3}$ & $\phi_{3} = \Phi_{3}-\pi$ &
    $\phi_{3} = \Phi_{3}$ &  $\phi_{3} = -\Phi_{3}$
\end{tabular}\end{center}
Thus, provided $\theta_{2}$ and $\theta_{3}$ can be found, a field can be 
generated by taking the subsequently found $\phi_{2}$ and $\phi_{3}$ and
creating the four corner vectors of the face.  These vectors are then
transformed through the transformation matrix $T$ to allow for more generality
in the direction of the field.  Following this, positive constants can multiply
the resulting corner vectors, so they have any required magnitude.  From
these four vectors, a trilinear field may be extrapolated, with the condition
\(\nabla.\vec{B} = 0\) satisfied, using
\begin{equation}
  \vec{B}(\vec{x}) = \left(\begin{array}{c} a_{1} \\ a_{2} \\ a_{3}
    \end{array}\right)+\left(\begin{array}{ccc} b_{1} & c_{1} & \alpha \\ a_{2}
    & c_{2} & \beta \\ a_{3} & c_{3} & -(b_{1}+c_{2})
    \end{array}\right)\left(\begin{array}{c} x \\ y \\ z
    \end{array}\right)+\left(\begin{array}{ccc} \varepsilon & \gamma & d_{1} \\
    -\gamma & \delta & d_{2} \\ -d_{1} & -d_{2} & d_{3} \end{array}\right)
    \left(\begin{array}{c} yz \\ xz \\ xy \end{array}\right),
\end{equation}
where $\alpha$, $\beta$, $\gamma$, $\delta$ and $\varepsilon$ are arbitrary
constants, and
\[\begin{array}{rcl@{\hspace{.5in}}rcl}
  \langle a_{1},a_{2},a_{3} \rangle & = & \vec{B}_{000}, &
  \langle b_{1},a_{2},b_{3} \rangle & = & \vec{B}_{100}-\vec{B}_{000}, \\
  \langle c_{1},a_{2},c_{3} \rangle & = & \vec{B}_{010}-\vec{B}_{000}, &
  \langle d_{1},a_{2},d_{3} \rangle & = & \vec{B}_{110}-\vec{B}_{100}
                                         -\vec{B}_{010}+\vec{B}_{000}.
\end{array}\]

\end{document}